\documentclass[mathleft
]{an}
\usepackage{graphicx}
\usepackage{times}
\usepackage{color}
\begin{document}

\Pagespan{1111}{}
\Yearpublication{2007}%
\Yearsubmission{2007}%
\Month{12}%
\Volume{328}%
\Issue{10}%
\DOI{10.1002/asna.200710865}%

\title{A coupled model of magnetic flux generation and transport in stars\thanks{Movies 
are available via http://www.aip.de/AN/movies}}

\author{E. I\c{s}\i k\thanks{Corresponding author:
  \email{ishik@mps.mpg.de}\newline}
\and D. Schmitt
\and  M. Sch\"ussler
}
\titlerunning{Magnetic flux generation and transport}
\authorrunning{E. I\c{s}\i k, D. Schmitt, M. Sch\"ussler}
\institute{
Max-Planck-Institut f\"ur Sonnensystemforschung, Max-Planck-Str. 2,
D-37191 Katlenburg-Lindau, Germany
}

\received{2007 Sep 10}
\accepted{2007 Oct 25}
\publonline{2007 Dec 15}


\abstract{%
  We present a combined model for magnetic field generation 
and transport in cool stars with outer convection zones. 
The mean toroidal magnetic field, which is generated by a cyclic thin-layer 
$\alpha\Omega$ dynamo at the bottom of the convection zone is taken to determine 
the emergence probability of 
magnetic flux tubes in the photosphere. Following the 
nonlinear rise of the unstable thin flux tubes, emergence latitudes and tilt angles 
of bipolar magnetic regions are determined. These quantities are put 
into a surface flux transport model, which simulates the surface evolution 
of magnetic flux under the effects of large-scale flows and turbulent 
diffusion. 
First results are discussed for the case of the Sun and for more rapidly 
rotating solar-type stars.}

\maketitle

\section{Introduction}
The increasing observational knowledge on stellar magnetic activity makes 
it possible to use the observed activity patterns to constrain 
stellar dynamo models (Strassmeier~\cite{kgs05}; Collier~Cameron~\cite{cc07}; 
\sloppy{Holzwarth, Mackay \& Jardine~\cite{vrh07}}). 
On the one hand, many solar and stellar dynamo models 
in the literature make the 
assumption that the toroidal magnetic fields created in stellar interiors represent 
also the surface emergence patterns. 
On the other hand, numerical simulations of the rise 
of flux tubes in the convection zone 
(Caligari, Moreno-Insertis, \& Sch\"ussler~\cite{cale95}) 
have been successful in reproducing the 
observed properties of sunspot groups, among which e.g., the tilt angle is  
of particular importance in surface flux transport models and 
Babcock-Leighton-type dynamos (Dikpati \& Charbonneau~\cite{dikpati99}). 
Therefore, it is now possible to consider in a consistent way the interrelations
between the dynamo 
mechanism operating in stellar interiors, transport of toroidal magnetic 
flux in the convection zone, and the emerging flux which evolves under the 
effects of surface flows. 
Our approach combines models for three processes: 
1) the dynamo operating in the overshoot layer at the bottom of the convection 
zone, 2) stability and rise of magnetic flux tubes through the convection zone, 
3) transport of magnetic flux on the surface. 

\section{The model}
\label{sec:model}
We consider stars with solar internal structure and 
differential rotation, 
keeping the difference between the minimum and 
maximum angular velocities, $\Delta\Omega$, independent of 
the stellar rotation rate, which is a plausible assumption for rapidly rotating 
stars (e.g., Barnes et al.~\cite{barnes}). The only stellar parameter 
that we vary is the surface rotation period at the equator, $P_{\rm rot}$. 

{\it The dynamo model}. The generation of 
magnetic flux is described by a dynamo model, which 
provides the toroidal component of the mean magnetic field. As a simple example, 
here we consider an $\alpha\Omega$ dynamo operating in a thin layer at the bottom 
of the convection zone (Schmitt~\&~Sch\"ussler~\cite{ss89}). 
We assume a Sun-like radial differential rotation profile, according to helioseismic 
observations (Schou et al. \cite{schou98}). 
A negative $\alpha$-effect is assumed for latitudes below $35^\circ$, where the 
radial shear is positive. 
This choice is motivated by an $\alpha$-effect due to flux tube instabilities
(Ferriz~Mas, Schmitt \& Sch\"ussler~\cite{afm94}) or unstable magnetostrophic 
waves (Schmitt~\cite{schmitt03}). The maximum strength of the 
$\alpha$-effect (at $\lambda$=$\pm 17.5^\circ$) is assumed to be proportional 
to the stellar equatorial rotation rate, $\Omega_0$. The turbulent diffusivity 
of $5.6\cdot 10^{11}$ cm$^2$~s$^{-1}$ is chosen such that the cycle period 
becomes 11 years in the case of the Sun. 

{\it The rise of flux tubes in the convection zone}. 
In this part of the model, the thin flux tube approximation is considered. 
Magnetic flux tubes are assumed to form (e.g., by magnetic Rayleigh-Taylor 
instability) within a layer of toroidal magnetic field created by the radial 
shear in the upper tachocline. 
When the field strength of a tube exceeds a threshold, 
the Parker (undulatory) instability sets in, and the flux tube forms one or 
two rising loops.
The probability for a flux tube 
to erupt from a given latitude $\lambda$ and at a given time $t$ is assumed to be 
proportional to the mean toroidal field, $B(\lambda,t)$, provided by the dynamo model. 
The total number of erupting flux tubes per activity cycle is scaled with $\Omega_0$, 
in accordance with the observed relation between rotation and activity 
(e.g., Montesinos~et~al.~\cite{montesinos01}). 
Considering the linear stability of flux tubes in the mid-overshoot 
region (Ferriz Mas \& Sch\"ussler~\cite{afmsch95}), 
unstable flux tubes near the stability limit are chosen. 
The corresponding field strengths are around $10^5$ G for the case of the Sun. 
Having determined the eruption times, initial latitudes, and the corresponding field 
strengths of the individual flux tubes, we perform simulations 
of their buoyant rise 
(cf. Caligari et al.~\cite{cale95}) 
in order to determine the emergence 
latitudes and the tilt angles of the resulting bipolar magnetic regions at the surface. 
The separation of large bipolar regions 
on the Sun corresponds to much higher azimuthal wavenumbers ($m=10-60$) than found 
for Parker-unstable flux tubes ($m=1-2$). 
%
A solution to this problem 
is not available yet. There are, however, indications that bipolar regions 
dynamically disconnect from their roots at depths of only a few Mm below the surface and 
at an early stage of emergence, and this limits the azimuthal separation 
of emerging bipoles (Sch\"ussler \& Rempel~\cite{schrmp05}). 

{\it Surface flux transport}. 
The emergence times, latitudes, and tilt angles resulting from the previous step 
are taken as the input (source term) for a surface flux transport model 
(Baumann et al.~\cite{bsss04}, \cite{bss06}). The source term defines 
bipolar magnetic regions as a function of position and time. 
The flux transport code simulates the evolution of purely vertical 
magnetic fields at the 
surface, under the effects of Sun-like latitudinal differential rotation, 
meridional flow, and supergranular diffusion.
The emergence longitudes of bipolar regions are 
assumed to be randomly distributed. The areas of the bipolar regions are determined 
at random with a number distribution $N(A)\sim A^{-2}$, which represents the 
observed distribution of solar bipolar magnetic regions (Schrijver \& Harvey~\cite{sh94}). 

\section{Results}
\label{sec:results}
\begin{figure}
   \centering
   \includegraphics[width=0.85\linewidth]{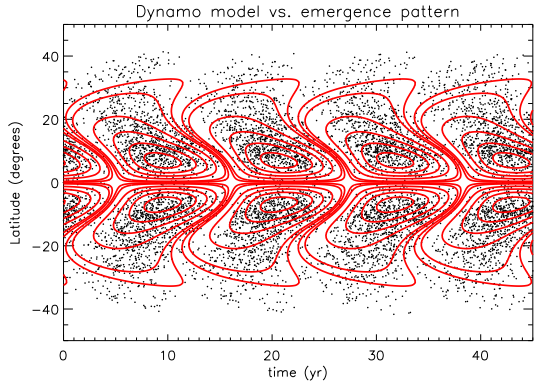}
   \caption{Time-latitude diagram of the dynamo-generated mean toroidal magnetic 
   field $B(\lambda,t)$ at the bottom of the convection zone (contours) and of 
   the flux loops emerging at the surface (dots) for a Sun-like star with 
   $P_{\rm rot}=27$ d. The dynamo waves and the emergence pattern closely match 
   in this case.}
   \label{fig:sun1}
\end{figure}

First we consider the Sun-like case with $P_{\rm rot}=27$ d. 
Figure~\ref{fig:sun1} shows the time-latitude diagram of the dynamo-generated 
toroidal magnetic field at the bottom of the convection zone
and the locations of flux emergence. The toroidal field contours almost coincide 
with the overall emergence pattern, because of the small poleward deflection 
of rising flux tubes for a slow rotator like the Sun. This result justifies 
an implicit assumption that is often made when interpreting solar dynamo models: 
the surface activity pattern reflects the dynamo wave pattern.  
\begin{figure}
   \centering
   \begin{minipage}{\columnwidth}
      \includegraphics[width=0.9\linewidth]{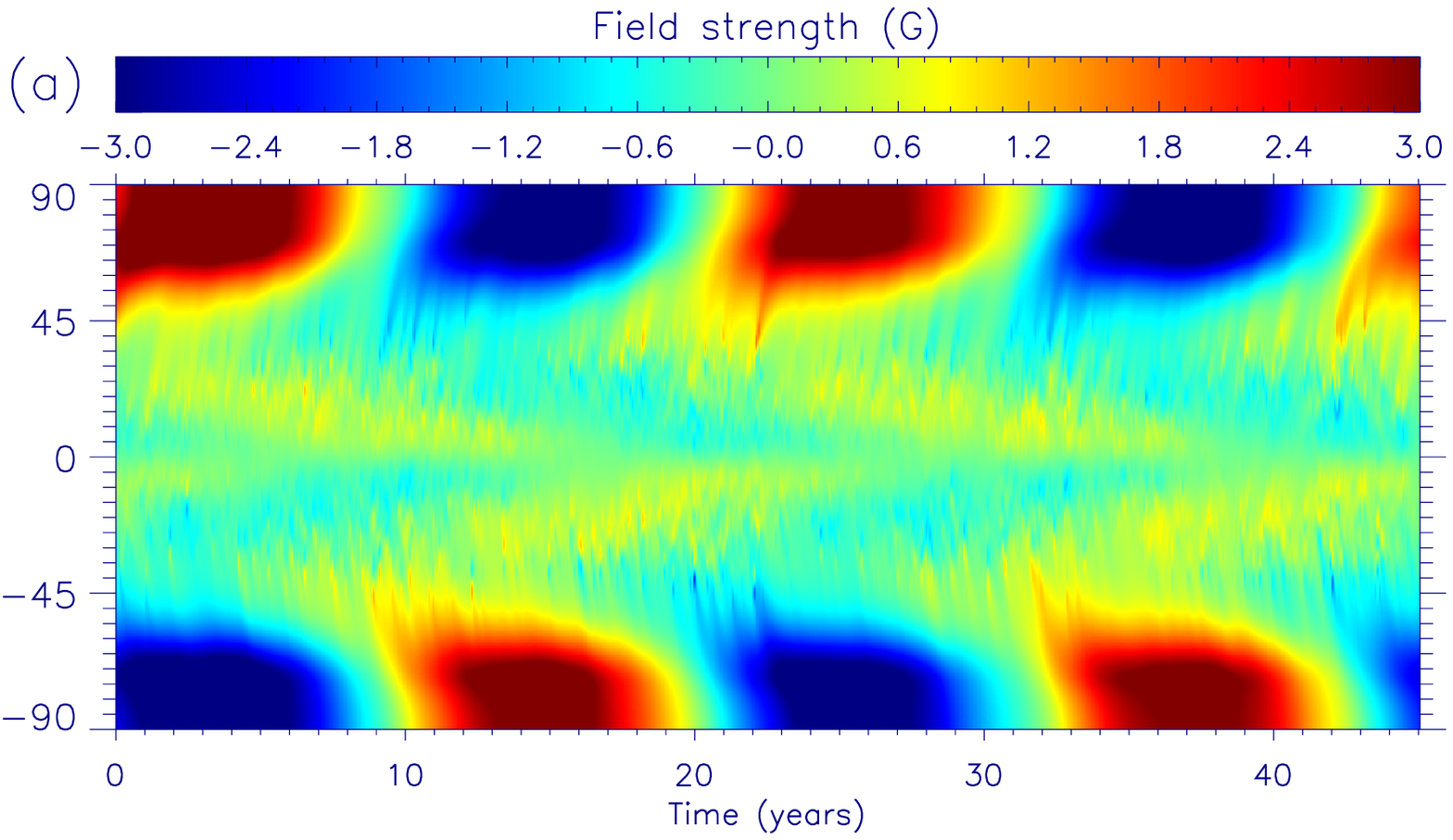} \\
      \includegraphics[width=0.9\linewidth]{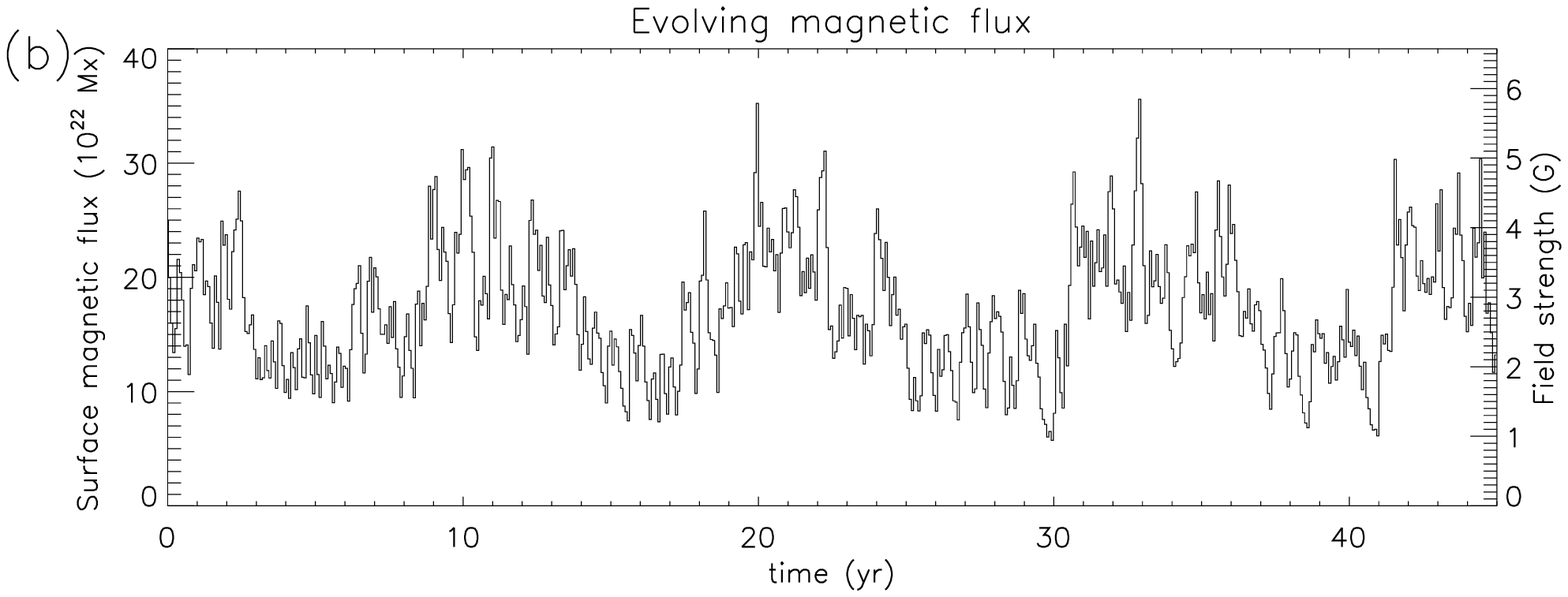}
   \end{minipage}
   \caption{(a) Time-latitude diagram of the azimuthally 
   averaged strength of the radial surface magnetic field for the solar model. 
   (b) Time variation of the 
   surface-integrated magnetic flux. The values are averaged over 27-day 
   time intervals.}
   \label{fig:sun2}
\end{figure}
Azimuthal averages of the radial surface magnetic field are shown in a 
time-latitude diagram in Fig.~\ref{fig:sun2}a and the time evolution of the 
total surface magnetic flux is given in Fig.~\ref{fig:sun2}b.
The values are close to the results of Baumann et al.~(\cite{bsss04}), 
with the exception that polar fields are slightly weaker in our case. 
The reason is that the tilt 
angles resulting from the flux tube rise are systematically smaller than 
for the functional latitude dependence assumed by Baumann et al.~(\cite{bsss04}), 
but actually provide a better match to the observations 
(see Fig.~12 in Caligari et al.~\cite{cale95}). 
An animation of the evolving surface flux is given in the supplementary file 
{\tt Bsurf\_27d.gif}. 

\begin{figure}
   \centering
   \includegraphics[width=0.8\linewidth]{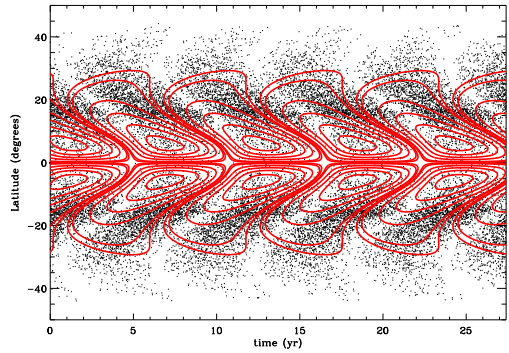}
   \caption{Same as Fig.~\ref{fig:sun1}, for $P_{\rm rot}=10$ d.}
   \label{fig:act1}
\end{figure}

We next consider a star with a rotation period of 10 d. 
The generated toroidal field and the emergence patterns are shown in 
Fig.~\ref{fig:act1}. 
In this case, the poleward deflection of rising flux tubes is stronger than in 
the case of the Sun. The responsible effect is the component of the Coriolis force, 
directed towards the rotation axis (resulting from a flow along the flux tube 
in order to conserve angular momentum), and is proportional to the rotation rate. 
\begin{figure}
   \centering
   \begin{minipage}{\columnwidth}
      \includegraphics[width=0.9\linewidth]{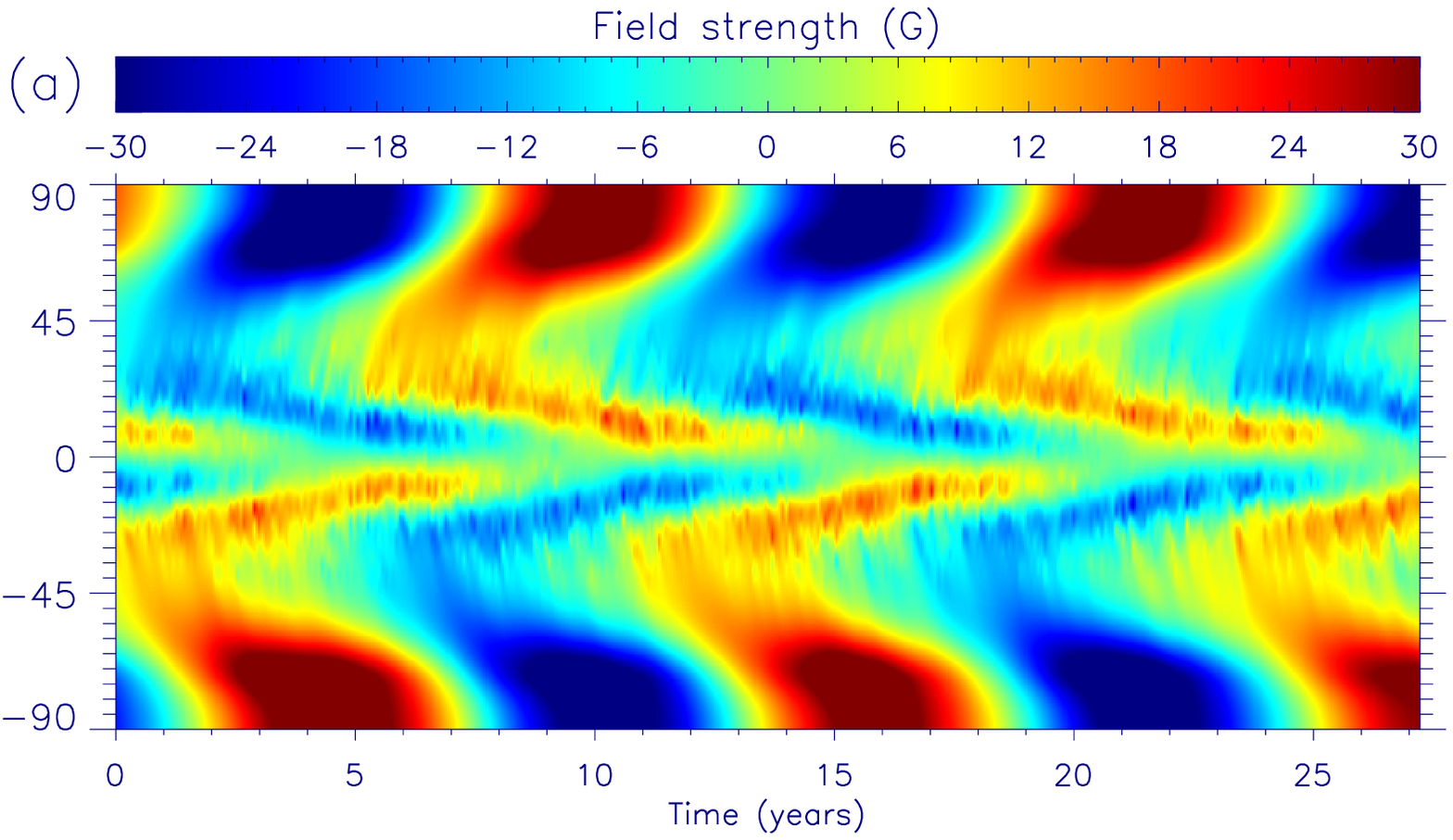} \\ 
      \includegraphics[width=0.9\linewidth]{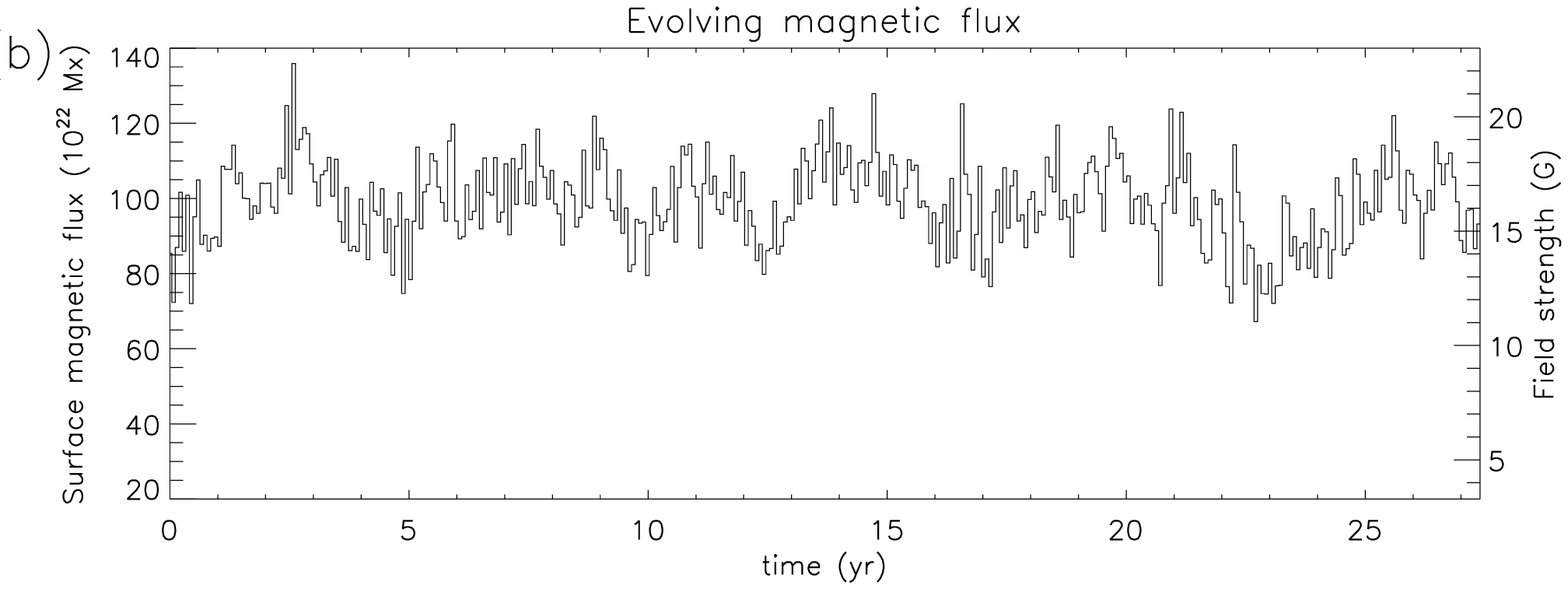}
   \end{minipage}
   \caption{Same as Fig.~\ref{fig:sun2}, for $P_{\rm rot}=10$ d.}
   \label{fig:act2}
\end{figure}
As shown in Fig.~\ref{fig:act2}a, we have strong polar fields (note the different 
saturation level of the colour table, as compared to 
Fig.~\ref{fig:sun2}a). This is due to 1) higher emergence rate, 2) larger tilt angles 
owing to faster rotation. 
An animation of the evolving surface flux is given in the file {\tt Bsurf\_10d.gif}. 
Figure~\ref{fig:act2}b shows that a cycle signal is no longer clearly observable 
when we consider the evolving surface magnetic flux. The reasons are 
1) a larger degree of overlap between consecutive cycles owing to a stronger 
$\alpha$-effect, and 2) strong polar magnetic fields in antiphase 
with the emerging flux.
\begin{figure}
   \centering
   \includegraphics[width=0.9\linewidth]{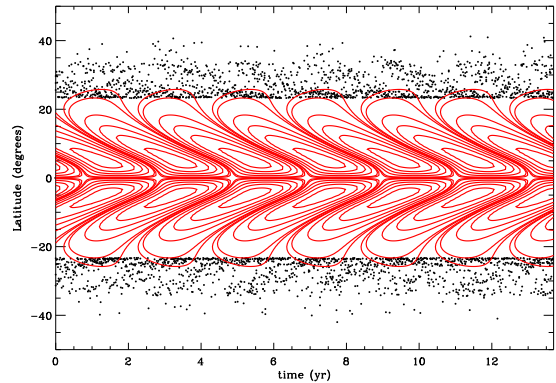}
   \caption{Same as Fig.~\ref{fig:act1}, for $P_{\rm rot}=2$ d. 
   The dots are shown with one-month intervals. The emergence 
   pattern is completely different in comparison with the dynamo wave.}
   \label{fig:vact}
\end{figure}

When we decrease the rotation period to 2 d, the poleward deflection of flux 
tubes causes 
a significant difference between the dynamo waves and the surface emergence pattern, as 
shown in Fig.~\ref{fig:vact}. An animation of surface transport is given in the 
supplementary file {\tt Bsurf\_2d.gif}. 
The tilt angles of emerging bipolar regions are much larger 
(around $35^\circ$) than in the solar case, leading to two latitudinal belts 
of opposite magnetic polarity. Meridional transport of the high-latitude belt leads 
to strong polar fields, with field strengths up to 30~G.

\section{Conclusions and outlook}
\label{sec:conc}
We have developed a consistently coupled model of magnetic field 
generation and transport in the Sun and other cool stars. Similar to 
Schrijver \& Title~(\cite{st01}) and I\c{s}\i k, Sch\"ussler \& Solanki~(\cite{iss07}), 
we find significant polar magnetic regions for rotation periods of 10~d and 2~d, 
in addition to low-latitude activity. 
For $P_{\rm rot}=10$~d, the surface flux transport blurs the periodic signal 
from the dynamo model. This indicates that for some stars the cyclic 
dynamo may be hidden in a non-cyclic surface activity, which is 
led by the combined effects of the dynamo (cycle overlap), large tilt angles, 
and the meridional transport. Furthermore, for 
$P_{\rm rot}=2$~d we have found that the dynamo wave pattern and the emerging surface 
flux are completely different, owing to the strong poleward deflection of rising 
flux tubes by the Coriolis force. 

As next steps, we plan to quantify the loss of magnetic flux from the dynamo layer 
in a consistent manner and consider it 
in the dynamo equations as a nonlinear term. Furthermore, we 
intend to develop a scheme which includes the evolving surface flux into a two 
dimensional flux transport dynamo as the source of the poloidal component at 
the bottom of the convection zone.

\end{document}